\documentclass{book}
\usepackage{qpt_book}
\usepackage{amsfonts,amsmath,amssymb} 
\usepackage{graphicx}
\usepackage{epstopdf}
\usepackage{wasysym}

\newcommand{\onlinecite}[1]{\cite{#1}}
\newcommand{\beq}{\begin{equation}}
\newcommand{\eeq}{\end{equation}}
\newcommand{\bea}{\begin{eqnarray}}
\newcommand{\eea}{\end{eqnarray}}
\newcommand\bftheta{\boldsymbol{\theta}}

\newcommand{\ket}[1]{\left|{#1}\right\rangle}
\newcommand{\bra}[1]{\left\langle{#1}\right|}

\def\openone{\leavevmode\hbox{\small1\kern-4.2pt\normalsize1}}
\def\A{\mathcal{A}}
\def\B{\mathcal{B}}

\def\K{\mathcal{K}}
\def\L{\mathcal{L}}

\def\S{\mathcal{S}}
\def\SvN{S_{\rm vN}}
\def\Stopo{S_{\rm topo}}
\def\Tr{\textrm{Tr}}

\begin{document}

\pagestyle{myheadings}


\markboth{Perspectives on Quantum Phase Transitions}
{Topological Order and Quantum Criticality}

\chapter{Topological Order and Quantum Criticality}

\vskip -0.5cm
{\bf \large Claudio Castelnovo} \\
{\it Rudolf Peierls Centre for Theoretical Physics and Worcester College,\\  Oxford University, Oxford OX1 3NP, UK} \\
\vskip 2mm
\noindent
{\bf \large Simon Trebst} \\
{\it Microsoft Research, Station Q, University of California,\\ Santa Barbara, California 93106, USA} \\
\vskip 2mm
\noindent
{\bf \large Matthias Troyer} \\
{\it Theoretische Physik, Eidgen\"ossische Technische Hochschule Zurich,\\ 8093 Zurich, Switzerland} \\
\\

\noindent
In this chapter we discuss aspects of the quantum critical behavior that
occurs at a quantum phase transition separating a topological phase
from a conventional one. 
We concentrate on a family of quantum lattice models, namely certain
deformations of the toric code model, that exhibit continuous quantum 
phase transitions. 
One such deformation leads to a Lorentz-invariant transition in the 
3D Ising universality class. 
An alternative deformation gives rise to a so-called conformal quantum 
critical point where equal-time correlations become conformally invariant 
and can be related to those of the 2D Ising model.
We study the behavior of several physical observables, such as non-local 
operators and entanglement entropies, that can be used to characterize 
these quantum phase transitions. 
Finally, we briefly consider the role of thermal fluctuations and related 
phase transitions, before closing with a short overview of field theoretical 
descriptions of these quantum critical points.


\section{Introduction}
\label{CTT_sec: intro}

Topological phases are distinct both from high-temperature disordered phases and
conventional ordered ones: they exhibit order, albeit of a type which cannot
be defined locally \cite{CTT_Wen1989}. 
They are often described by an emergent symmetry that is
captured by a new set of quantum numbers such as a ground-state degeneracy and
fractional quasiparticle statistics.
The archetypal physical realizations of topologically ordered states are 
fractional quantum Hall states.

This chapter is concerned with topological order and quantum critical behavior in the
context of microscopic lattice models. Most of our discussion will concentrate
on {\em time-reversal invariant} systems which in more general terms can be described 
by so-called quantum double models and are thus quite distinct from fractional quantum Hall 
states. 
We start by shortly reviewing the {\em toric code} model, a particularly simple 
and exactly solvable spin-1/2 model that exhibits an Abelian topological phase. 
Our subsequent discussion of quantum phase transitions involving topologically ordered 
phases of matter is structured around two types of deformations of the toric code which we 
dub ``Hamiltonian deformation"  and ``wavefunction deformation", respectively, and that give 
rise to distinct quantum critical points.
We then turn to finite-temperature phase transitions and discuss under which circumstances
topological order is robust with regard to thermal fluctuations.

%

\subsection{The toric code}
\label{CTT_sec:ToricCode}

\begin{figure}[t]
         \begin{center}
	\includegraphics[width=0.9\columnwidth]{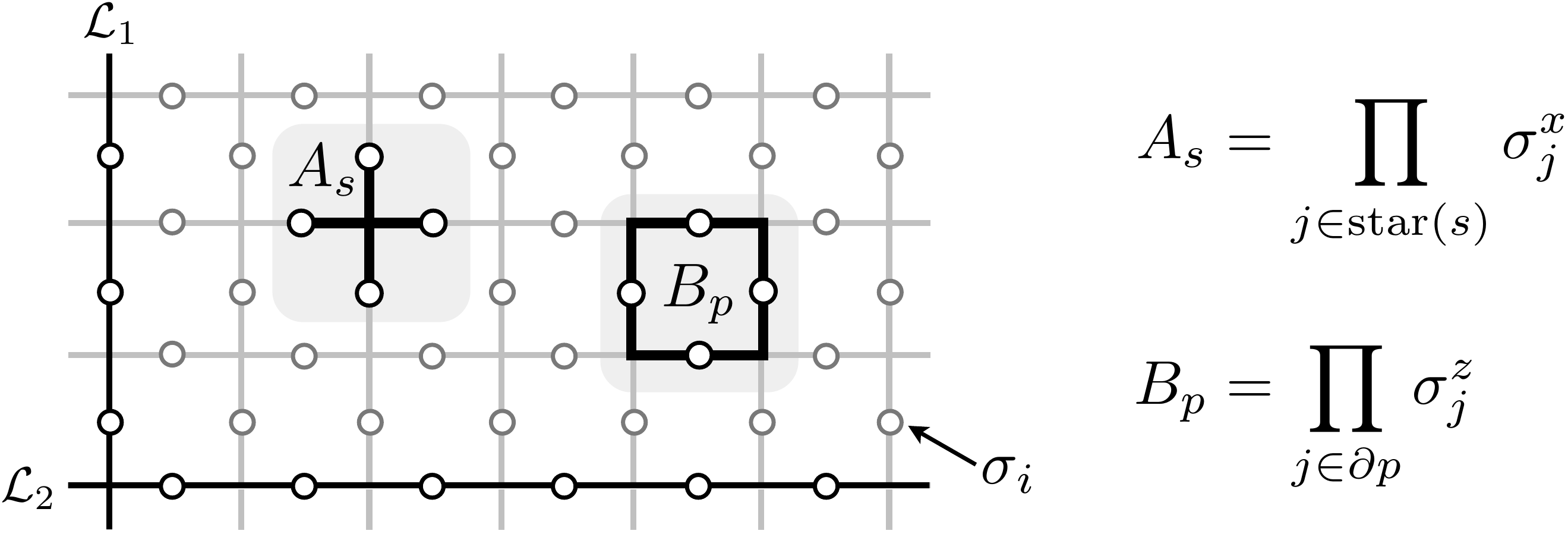}
	\end{center}
	\caption{{\em The toric code model:}
	                The elementary degrees of freedom are SU(2) spin-1/2's on the bonds 
	                of a square lattice. The star operator $A_s$ acts on the
	                spins surrounding a vertex and the plaquette operator $B_p$ 
	                acts on the spins surrounding a plaquette.
			}
	\label{CTT_Fig:ToricCodeModel}
\end{figure}

The toric code (TC)  \cite{CTT_Kitaev2003} is an exactly solvable model of SU(2) spin-1/2 degrees of freedom,
which is typically defined on a square lattice, but can be adapted to other two-dimensional lattice geometries 
(and even higher-dimensional lattices).
The Hamiltonian is generally defined as a sum of vertex and plaquette terms
\begin{equation}
   \mathcal{H}_{\rm TC} = -J_e \sum_s A_s - J_m \sum_p B_p \,,
   \label{CTT_Eq:ToricCode}
\end{equation}
where the star operator $A_s$ acts on all spins adjacent to a vertex $s$ by coupling their
$\sigma^x$ spin components, $A_s = \prod_{j \in {\rm star}(s)} \sigma_j^x$,
and the plaquette operator $B_p$ acts on all spins surrounding a plaquette $p$ 
by coupling their $\sigma^z$ spin components, $B_p = \prod_{j \in \partial p} \sigma_j^z$. 
We typically consider the case $J_e, J_m > 0$.
In the case of the square lattice, the star and plaquette operators are both 4-spin operators
which are Hermitian and have eigenvalues $\pm 1$.

The toric code model can be solved exactly, since all terms in the Hamiltonian commute with 
one another. Indeed, $A_s B_p = B_p A_s$, $\forall\,s,p$ since any given pair of vertex and plaquette operators either share 0 or 2 bonds, and thus all minus
signs arising from the commutation of $\sigma^x$ and $\sigma^z$ on those bonds cancel.
This allows to construct eigenstates $\ket{\xi}$ of the toric code as common eigenstates of 
all terms in the Hamiltonian.
For a ground state the minimization of the energy then gives the ``stabilizer conditions"
$A_s \ket{\xi} = B_p \ket{\xi} = \ket{\xi}$ for each vertex/plaquette term, 
and an overall ground-state energy $E_0 = -N (J_e + J_m)$, 
where $N$ is the number of sites on the lattice. 

For a square lattice with periodic boundary conditions in both space directions (thus forming a torus) 
the sign of the vertex (plaquette) terms can only be flipped on an {\em even} number of vertices
(plaquettes). 
This implies that there are two overall constraints
\[
   \prod_s A_s = \prod_p B_p = +1 \,.
\]
This in turn lets us estimate the ground-state degeneracy on a torus: 
The total Hilbert space has $2^{2N}$ states. 
With the above constraint there are only $2N-2$ independent choices of the $A_s$ and $B_p$ 
operators,  thus imposing $2^{2N-2}$ conditions. 
As a consequence, we have $2^{2N} / 2^{2N-2} =  4$ ground states on the torus.

Another way to see how this ground-state degeneracy arises is by explicitly constructing the
ground-state wavefunctions. In doing so we solve the toric code Hamiltonian in the  $\sigma^z$ basis 
and introduce classical variables $z_j = \pm 1$ to label the $\sigma^z$ basis states. 
This allows to define a plaquette flux 
\[ 
    \phi_p(\bold{s}) = \prod_{j \in \partial p} z_j
\]
for each classical spin configuration $\bold{s}=\{z_j\}^{2N}_{j=1}$. If $\phi_p = -1$, we say there is a vortex on plaquette $p$.
For a ground state $\ket{\psi}$ of the toric code we need to maximize each plaquette term 
$B_p \ket{\psi} = \ket{\psi}$ and thus find that a ground state contains no vortices
\[
     \ket{\psi} = \sum_{\{\bold{s}\,:\, \phi_p(\bold{s})=+1 \,\, \forall p\}} c_{\bold{s}} \ket{\bold{s}} \,.
\label{CTT_eq: toric code GS}
\]
To maximize the vertex operator, e.g. $A_s \ket{\psi} = \ket{\psi}$,  all coefficients $c_{\bold{s}}$ are then
required to be equal (for each orbit of the action of the star operators). 
A ground state of the toric code is thus an equal-weight superposition of vortex-free spin configurations.

On the square lattice there are four distinct ways to construct such equal-weight superpositions, 
giving rise to a four-fold ground-state degeneracy. This can be seen by considering the following
function measuring the flux through an extended loop $\L$  (similar to a Wilson loop), 
as indicated in Fig.~\ref{CTT_Fig:ToricCodeModel}
\[
    \Phi_{\L} (\bold{s}) = \prod_{j \in \L} z_j \,.
\]
If the loop $\L$ is a contractible loop, then the flux $\Phi_{\L}$ is simply the product of the 
plaquette fluxes $\phi_p$ within the loop (reminiscent of Stoke's theorem), 
and for each ground state $\Phi_{\L} = +1$.
If on the other hand the loop $\L$ is an essential loop on the torus (which cannot be contracted),
then the flux $\Phi_{\L} = \pm 1$ defines a conserved quantity, since an arbitrary 
star operator $A_s$ overlaps with the loop $\L$ in either 0 or 2 bonds and thus preserves $\Phi_{\L}$.
For the torus there are two independent, essential loops $\L_1$, $\L_2$ wrapping the torus as 
illustrated in Fig.~\ref{CTT_Fig:ToricCodeModel}.
The ground states of the toric code can then be labeled by two conserved quantities 
$\Phi_{\L_1} = \pm 1$ and $\Phi_{\L_2} = \pm 1$. The resulting degeneracy originates from the 
four distinct {\em topological sectors} defined by the essential loops.
Embedding the toric code onto the surface of a more general manifold, one finds that the 
ground-state degeneracy increases as $4^g$ with the genus $g$ of the underlying surface. 

We note in passing that one can find another description of the eigenstates of the 
toric code by solving the Hamiltonian in the $\sigma^x$ basis.
Labeling the  basis states by classical variables $x_j = \pm 1$ and identifying the 
$x_j = +1$ states with loop segments on the corresponding bonds, one finds that the ground states of the toric code form a 
{\em quantum loop gas} with fugacity $d=1$. The four topological sectors then correspond to loop
configurations with an even or odd number of loops wrapping the torus across two independent, 
non-contractible cuts.

The toric code Hamiltonian can also be viewed as a particular lattice regularization of an Ising gauge theory
\cite{CTT_Kogut1979}. 
In this language the spins residing on the bonds correspond to $Z_2$ valued gauge 
potentials, the star operators $A_s$ become gauge transformations, and their commutation with the 
plaquette flux operators $B_p$ implies an overall gauge invariance. 
Owing to this equivalence to an Ising gauge theory, the two distinct excitations of the toric code 
that arise from violating one of the stabilizer conditions of the ground states, 
$A_s \ket{\psi} = \ket{\psi}$ and $B_p \ket{\psi} = \ket{\psi}$, are commonly identified 
as {\em electric charges} and {\em magnetic vortices}, respectively.

For a given ground state a pair of electric charges is created by applying the $\sigma^z$-operator to a spin, 
which leads to a violation of the stabilizer condition $A_s \ket{\psi} = \ket{\psi}$ on the two adjacent 
vertices. Therefore, the energy cost to create two electric charges is $4J_e$. 
While a pair of electric charges does not need to reside on neighboring vertices, it remains 
connected for any separation by an electric path operator $\prod_{j \in \ell} \sigma^z_j$, where $\ell$ is a 
path connecting the two vertices. 
\iffalse
Similarly, a pair of magnetic vortices can be created by applying the $\sigma^x$-operator to a spin 
which results in a violation of the stabilizer condition $B_p \ket{\psi} = \ket{\psi}$ on the two adjacent 
plaquettes. The energy cost to create two magnetic vortices thus becomes $4J_m$. 
Again for arbitrary separation a pair of magnetic vortices on plaquettes $p_1$ and $p_2$ remains
connected by a magnetic path operator $\prod_{j \in \ell^*} \sigma^x_j$, where $\ell^*$ is a path on the dual 
lattice linking the plaquettes $p_1$ and $p_2$.
\else
Similarly for a pair of magnetic vortices, upon replacing $\ell$ with a path on the dual lattice. 
\fi

Both types of excitations are massive quasiparticles, and for the unperturbed toric code
Hamiltonian \eqref{CTT_Eq:ToricCode} they are static excitations which do not disperse. 
While single electric charges or magnetic vortices are bosons 
(considering the exchange with the same particle type), 
the composite quasiparticle of an electric charge plus a magnetic vortex constitutes 
a fermion. Moreover, electric charges and magnetic vortices exhibit an unusual, {\em mutual} exchange 
statistics in that they are mutual {\em semions} -- the simplest incarnation of Abelian anyons: 
when moving an electric charge around a magnetic vortex the wavefunction picks up a phase of 
$e^{i\pi} = -1$. 
This can best be seen when considering a wavefunction $\ket{\xi}$ of a state containing a single
vortex at a plaquette $\tilde{p}$, e.g. $B_{\tilde{p}}\ket{\xi} = -\ket{\xi}$. 
If we move an electric charge along a contractible loop $\L$ around this plaquette $\tilde{p}$, 
the wavefunction transforms as 
\[
    \ket{\xi} \; \to \; \prod_{j \in \L} \sigma^z_j \ket{\xi} \; = \prod_{p \; {\rm inside } \; \L} B_p \ket{\xi} = - \ket{\xi} \,,
\] 
which reveals the mutual semionic statistics.

In summary, the toric code is one of the simplest and most accessible quantum lattice models that gives rise 
to a quantum state with {\em topological order}. This exotic quantum order reveals itself in a (robust)
ground-state degeneracy, and massive, deconfined quasiparticles with mutual anyonic statistics.

%
%
\section{Quantum Phase Transitions}
\label{CTT_sec: QPTs}

While conventional ordered phases arise from the spontaneous breaking of a symmetry,
the converse is true for topological phases which exhibit an enhanced symmetry at low
energies. It is this emerging topological invariance which isolates the low-energy degrees 
of freedom from local perturbations and gives rise to stable topological phases.
However, even a local perturbation, if sufficiently strong, can eventually destabilize 
a topological phase and give rise to a quantum phase transition.

To elucidate the quantum critical behavior at such a transition we consider the pedagogical
example of the toric code in a magnetic field. A sufficiently large magnetic field aligns
all the spins with the field resulting in a paramagnetically ordered state, and thus destroys any 
topological order. 
The quantum phase transition that connects these two phases can be either first-order or
continuous depending on the way we couple the magnetic field to the local spin degrees 
of freedom in the toric code Hamiltonian.
We mostly focus on the latter case of a continuous transition and discuss two distinct 
types of quantum critical points that can occur.

First, we consider a magnetic field term $\vec{h} \cdot \vec{\sigma}$, where the magnetic
field $\vec{h}$ has no tranverse $h_y$ component 
\cite{CTT_Trebst2007,CTT_Tupitsyn2008,CTT_Vidal2008}, 
i.e., $\vec{h} \cdot \vec{\sigma} = h_x \sigma^x + h_z \sigma^z$.
This ``Hamiltonian deformation" gives rise to a quantum
critical point at which the system becomes scale invariant and {\em local} two-point correlation 
functions show a divergent behavior. This not only allows us to define a dynamical critical exponent, 
which is found to be $z=1$, but also enables us to characterize this Lorentz-invariant transition out of the
topological phase in terms of a conventional universality class, which for this transition turns out to be 
the classical 3D Ising model.

\begin{figure}[t]
         \begin{center}
	\includegraphics[width=\columnwidth]{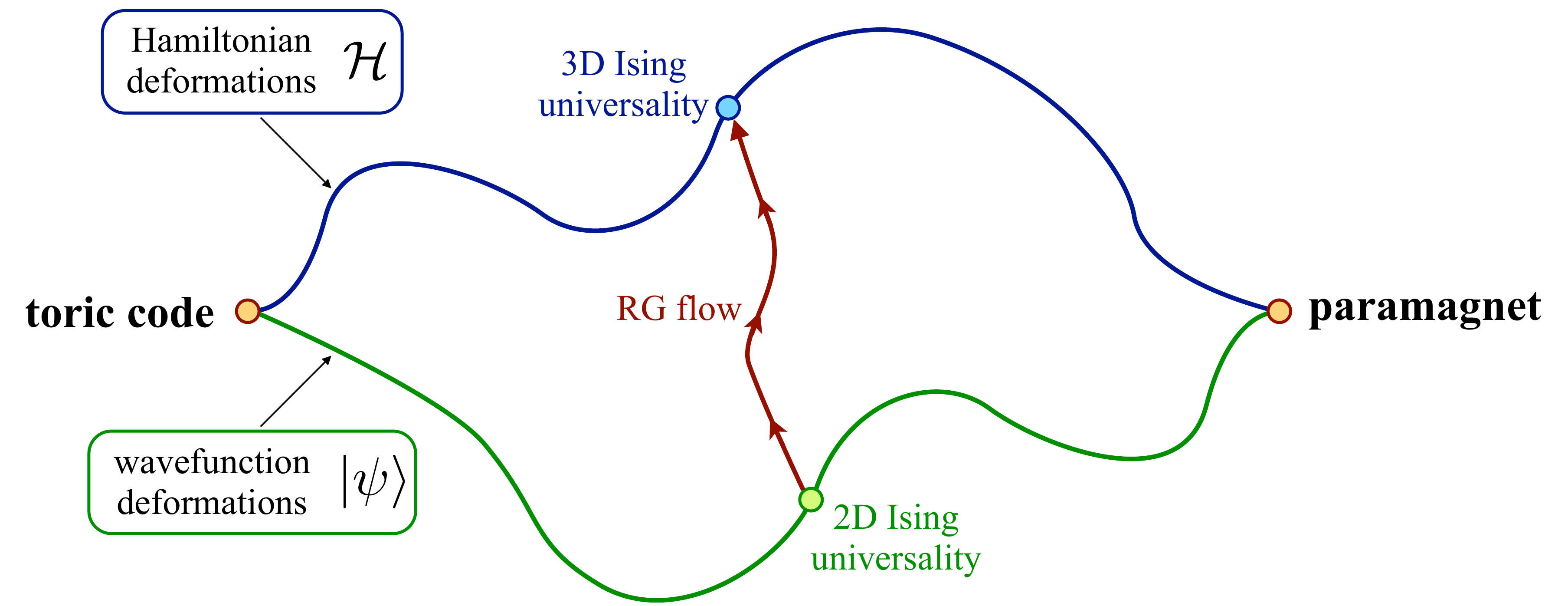}
         \end{center}
	\caption{{\em Toric code deformations:}
	                Two distinct types of quantum critical points are observed in phase space depending 
	                on the way the magnetic field is coupled to the local spin degrees of freedom in the
	                toric code.
	                }
	\label{CTT_Fig:ToricCodeDeformations}
\end{figure}

Second, we consider a magnetic field term $\exp{\left(- h_z  \sigma^{z}\right)}$, 
which for large field strength projects onto the fully polarized state, while allowing to
keep track of the ground-state wavefunction for all field strengths
\cite{CTT_Castelnovo2008_fK,CTT_Papanikolaou2007}.
We call this second scenario a ``wavefunction deformation". 
It gives rise to a so-called {\em conformal quantum critical point} \cite{CTT_Ardonne2004}, where
a peculiar type of dimensionality reduction is at play which allows to describe the equal-time correlations of
the ground state by a 2+0 dimensional conformal field theory which at this quantum critical point  
turns out to be the 2D Ising theory.
In contrast to the ``Hamiltonian deformation" the dynamical critical exponent is $z \neq 1$ for this 
second quantum critical point.
Despite the obvious differences in the intermediate quantum critical behavior both ``deformations" 
connect the same extremal ground states, namely those of the toric code with a fully polarized, 
paramagnetic state. 

We note in passing that (in the continuum limit) these two quantum critical points are connected
by a renormalization group flow from the conformal quantum critical point to the Lorentz-invariant
quantum critical point.
A schematic overview of these two types of deformations and their respective quantum
critical points is given in Fig.~\ref{CTT_Fig:ToricCodeDeformations}. 

Finally, we close by discussing the role of thermal fluctuations at the quantum critical
points of the deformed toric code models. This also allows to shed some light on the
question whether topological order can survive up to a finite temperature.

%
%

\subsection{Lorentz-invariant transitions} 
\label{CTT_sec: longitudinal field}

We first turn to the case of Lorentz-invariant quantum critical points which can be observed
when adding a longitudinal magnetic field to the toric code
\begin{equation}
   \mathcal{H}_{\rm TC+LMF} = -J_e \sum_s A_s - J_m \sum_p B_p 
                    				+ \sum_i \left( h_x \sigma^x_i + h_z \sigma^z_i \right) \,.
   \label{CTT_Eq:ToricCodeField}
\end{equation}
Whereas the model is no longer exactly solvable, we can readily understand the extremal cases of weak 
and strong magnetic fields. 

For $h_z=0$ the effect of a small field $h_x \ll J_e, J_m$ is the virtual creation and 
annihilation of pairs of magnetic vortices as discussed in the previous section, or the hopping
of a vortex excitation by one plaquette which allows the vortices to acquire a dispersion. 
In first-order perturbation theory this gives the usual tight-binding form in momentum space 
\[
     E(q_x, q_y) = 2J_m - 2h_x \left( \cos{q_x} + \cos{q_y} \right) + O( h_{x}^2 ) \,,
\]
which can be further expanded to higher orders \cite{CTT_Vidal2008}. 
A similar observation holds for the dual charge excitations.

\begin{figure}[t]
         \begin{center}
	\includegraphics[width=0.8\columnwidth]{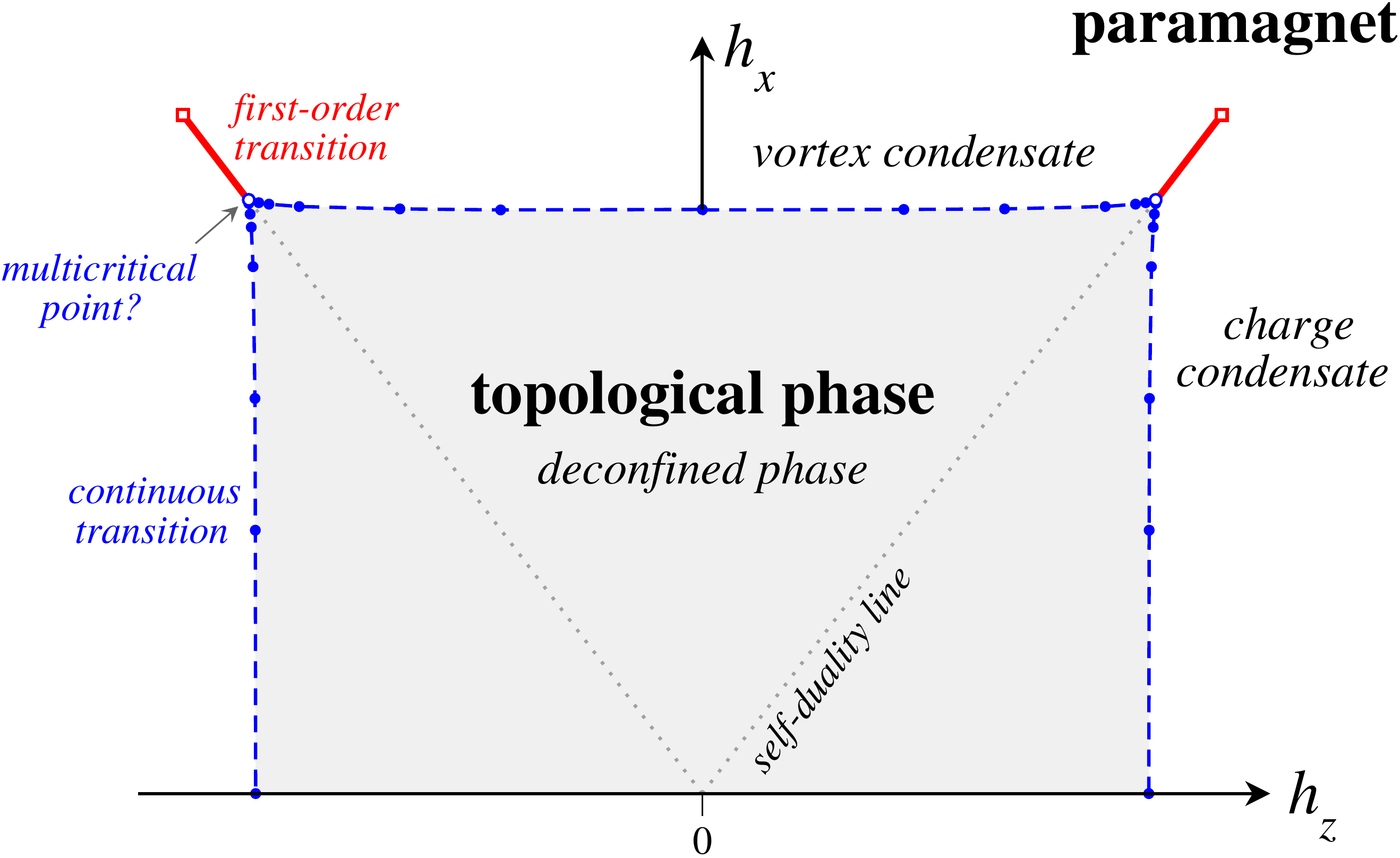}
	\end{center}
	\caption{{\em Phase diagram of the toric code in a longitudinal magnetic field:}
	                The dashed lines indicate continuous quantum phase transitions from
	                the topological phase (shaded area) to a paramagnetically ordered state.
	                The solid lines indicate first-order transitions.
	                The figure has been adapted from the numerical data of Ref.~\cite{CTT_Tupitsyn2008}
	                for the closely related classical, anisotropic $Z_2$ gauge Higgs model 
	                on a cubic lattice.}
	\label{CTT_Fig:ToricCodePhaseDiagram}
\end{figure}

With increasing magnetic field strength the quasiparticle gap eventually closes and the transition 
into the paramagnetic state can be understood as the Bose condensation of these quasiparticles. 
In the language of the Ising gauge theory this leaves us with two seemingly distinct possibilities,
as first discussed by Fradkin and Shenker \cite{CTT_Fradkin1979}:
A Higgs transition into a ``charge condensate" with concurrent vortex confinement for 
$h_x \ll h_z \approx J_e, J_m$, 
or by duality a confinement transition of the electric charges with the concurrent formation of
a ``vortex condensate" 
for $h_z \ll h_x \approx J_e, J_m$.
Both transitions gives rise to a {\em line of continuous transitions},
as illustrated schematically in Fig.~\ref{CTT_Fig:ToricCodePhaseDiagram},
where fluctuations of the system can be described in terms of two complimentary $Z_2$ 
order parameters, namely the amplitudes of the corresponding charge and vortex condensates.
As such the two transitions are clearly distinct and owing to the non-trivial mutual statistics 
of charges and vortices, these two lines of continuous transitions are not expected to join 
smoothly in the phase diagram.
On the basis of extensive numerical simulations it has recently been conjectured \cite{CTT_Tupitsyn2008,CTT_Vidal2008} 
that they meet at a {\em multicritical} point where they join with a first-order line separating the charge 
and vortex condensates for intermediate $h_x = h_z \approx J_e, J_m$.
This first-order transition is accompanied by a sudden change in the density of charges/vortices,
similar to a liquid-gas transition.
However, if we consider the paramagnetic phase in the large magnetic field limit of the original spin model, 
then we can continuously rotate the magnetic field in the $(h_x, h_z)$-plane without inducing a phase
transition. As a consequence, the charge and vortex condensates are actually the same phase, and they are 
connected in the large magnetic field limit. Indeed, the first-order line has a critical endpoint whose 
location has recently been determined by numerical simulations \cite{CTT_Tupitsyn2008,CTT_Vidal2008}.

To discuss the quantum critical behavior at one of the continuous transitions in the phase diagram
of Fig.~\ref{CTT_Fig:ToricCodePhaseDiagram} in more detail, we consider the case of a single-component
magnetic field in the $x$-direction, $h_x \neq 0$ and $h_{y,z}=0$.
As discussed before such a magnetic field will create/annihilate virtual vortex pairs that can hop through
the lattice, while charge excitations remain gapped and static. 
To describe the vortex dynamics in this limit we 
introduce a plaquette spin operator $\mu_p$ with eigenvalues $\mu_p^z = \pm 1/2$ depending on
the eigenvalue of the plaquette operator $B_p$ on the respective plaquette, e.g. $B_p = 2 \mu^z_p$. 
We can then rewrite the spin operators $\sigma^x$ in terms of these plaquette operators
as $\sigma^x_i = \mu_p^x \mu_q^x$, where $p$ and $q$ are the plaquettes separated by the bond $i$.
With these transformations in place, we can now recast the charge-free low-energy sector of the toric code 
in a single-component magnetic field in the form
\begin{equation}
   \mathcal{\tilde{H}}_{\rm TC+LMF} = - 2J_m \sum_p \mu^z_p 
   						+ h_x  \sum_{\langle p,q \rangle} \mu_p^x \mu_q^x 
   \label{CTT_Eq:TransverseFieldIsing}
\end{equation}
of the transverse field Ising model. Its well-known quantum phase transition 
thus describes the transition from the topological phase of the toric code 
into a paramagnetic state (or vortex condensate in the language of the Ising gauge theory). 
This continuous transition in the 3D Ising universality class occurs for coupling strength 
$(h_x / J_m)_c \approx 0.656 92(2)$~\cite{CTT_Bloete2002}, indicating that the topological
phase is remarkably stable -- it takes an $O(1)$ perturbation (in terms of the coupling strengths)
to destroy the topologically ordered state, similar to the case of many conventionally ordered states.

\begin{figure}[t]
         \begin{center}
	\includegraphics[width=0.8\columnwidth]{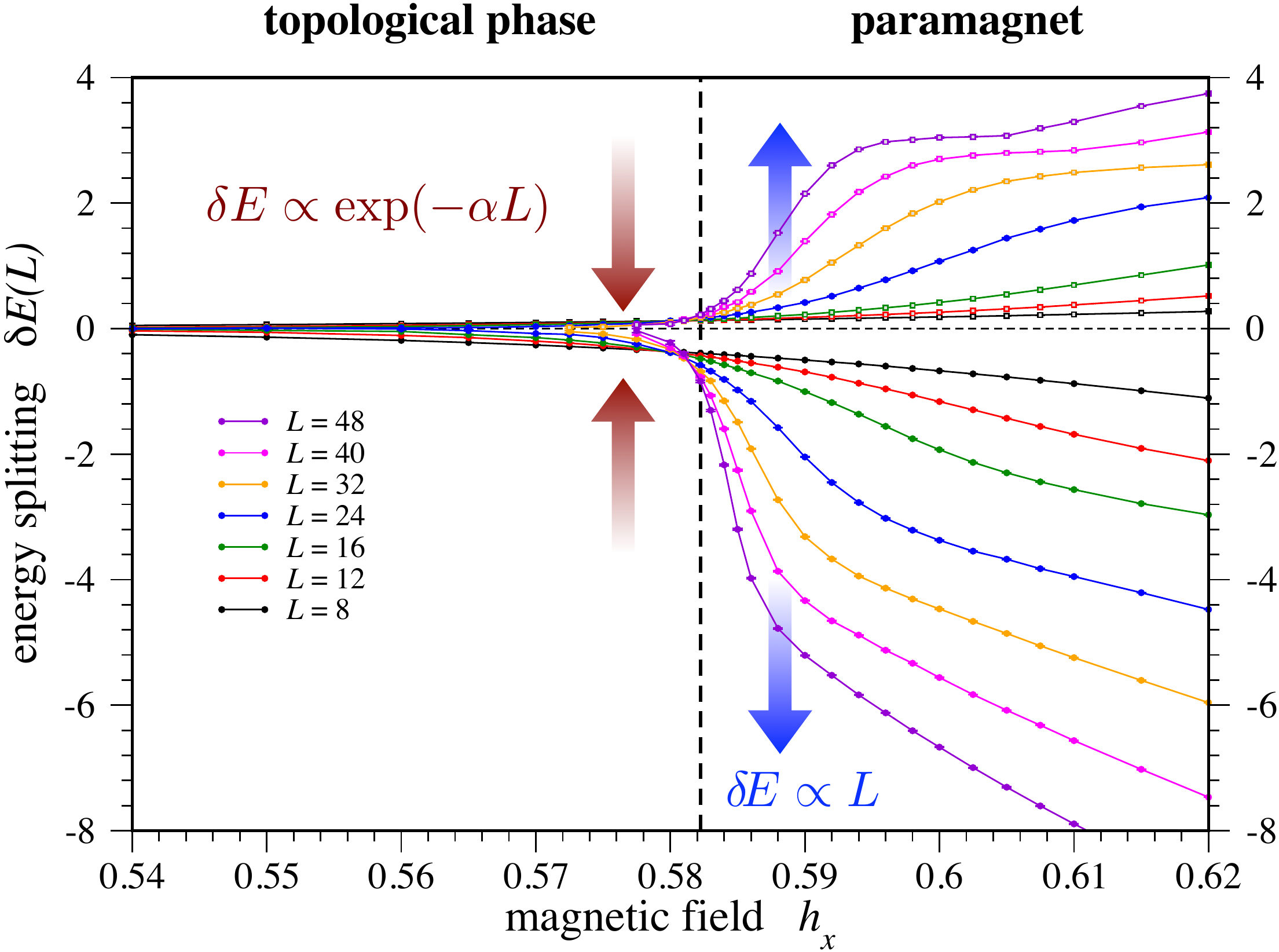}
         \end{center}
	\caption{{\em Splitting of the topological degeneracy:} 
		       The finite-size splitting of the ground-state degeneracy changes from exponential suppression
		       in the topological phase to linear splitting in the paramagnetic phase. Data for different linear
		       system sizes $L$ are plotted. 
		       (The critical coupling is slightly lower than stated in the text, as the numerical
		        simulations were done for discretized imaginary time and isotropic couplings in space-time.)
		      }
	\label{CTT_Fig:DegeneracySplitting}
\end{figure}

The mapping to the transverse field Ising model also allows to perform extensive Monte Carlo
simulations of this transition, and to keep track of the 
ground-state degeneracy across the quantum phase transition \cite{CTT_Trebst2007}. 
As shown in Fig.~\ref{CTT_Fig:DegeneracySplitting} the finite-size splitting of the degeneracy
changes from exponentially small in the topological phase to power-law 
in the paramagnetic phase (as a function of the linear system size $L$).

Returning to the full Hamiltonian \eqref{CTT_Eq:ToricCodeField} we can ask what effect a small magnetic 
field $h_z$ has on the quantum phase transition discussed above. Since such a field primarily 
induces dynamics in the gapped charge sector, the transition into the vortex condensate remains 
largely untouched, and we expect to observe a {\em line} of continuous transitions 
emanating from the single-component field limit discussed above.
In fact, one can map out the full phase diagram of the two-component case as shown in 
Fig.~\ref{CTT_Fig:ToricCodePhaseDiagram}. 
This is done through a sequence of transformations that allow to recast the toric code in a 
two-component magnetic field onto a classical, anisotropic $Z_2$ gauge Higgs model on a 
three-dimensional cubic lattice, as described in detail in Ref.~\cite{CTT_Tupitsyn2008}.

\subsubsection{Other Hamiltonian deformations}
\label{sec:otherdefomations}

A number of other Hamiltonian deformations of the toric code have been considered in the literature. 
Of particular interest is the case of a transverse field $h_y \sigma^y$. Contrary to the
longitudinal case, it is found to drive a first-order phase transition into a paramagnetic 
state \cite{CTT_Vidal2009}. 

The effect of dissipation was also studied \cite{CTT_Trebst2007} by coupling the toric code to an Ohmic
heat bath. A Thouless-type phase transition was found when the bath couples to the plaquette spins
$\mu_p^z$ such that it stabilizes the ``classical" state of the system.

%
%

\subsection{Conformal quantum critical points} 
\label{CTT_sec: modified longitudinal field}

We now turn to the second scenario of perturbing the toric code model via a  
{\em wavefunction deformation} that again connects the ground states 
of the toric code with a fully polarized, paramagnetic state
\cite{CTT_Castelnovo2008_fK,CTT_Papanikolaou2007}. 
Along this wavefunction deformation the system goes through a continuous
quantum phase transition which is distinct from the one we encountered
for the Hamiltonian deformations in the previous section. 
It belongs in fact to the family of {\em conformal quantum critical points}, 
first discussed by Ardonne, Fendley and Fradkin \cite{CTT_Ardonne2004}, 
where it is the ground-state wavefunction that becomes {\em scale invariant}, 
in a peculiar instance of dimensionality reduction. 
This in turn allows to describe the {\em equal-time} correlations in the ground-state
wavefunction of a two-dimensional quantum system by those of the related,
conformally invariant, {\em classical} system in $2$ dimensions -- contrary to the 
usual correspondence of a $D$-dimensional quantum system to a $D+1$ 
dimensional classical system.
This dimensionality reduction at a conformal quantum critical point is, of course, 
reminiscent of the well-known Rokhsar-Kivelson point of the 2D quantum 
dimer model~\cite{CTT_Rokhsar1988,CTT_Henley2004} 
where in a similar fashion equal-time correlations are captured by a ground-state
wavefunction that is the equal-weight superposition of all classical 2D dimer configurations.
Following earlier work by Henley \cite{CTT_Henley1997}, Ardonne {\em et al.} 
further argued that at a continuous transition, the conformal invariance of the 
ground state wave function enforces a quantum Lifshitz field theory 
description, with a characteristic $z=2$ dynamical exponent~\cite{CTT_Ardonne2004}.

To study wavefunction deformations of the toric code we employ a constructive 
technique dubbed ``stochastic matrix form" decomposition \cite{CTT_Castelnovo2005}
that allows to obtain generalized Rokhsar-Kivelson type Hamiltonians. 
With this approach we can derive an explicit expression for the ground state 
wavefunction along the deformation, and we can analyze the intervening conformal quantum 
critical point in great detail. 
In particular, we show that the quantum critical behavior is associated with a divergent
local length scale, despite the fact that there is no concomitant symmetry breaking 
phase transition described by a local order parameter. Finally,
we turn to entanglement measures and, after a short introduction, we discuss the 
behavior of the ``topological entropy" across the phase transition.


\subsubsection{Microscopic model for wavefunction deformation} 
\label{CTT_sec: model}

To implement the wavefunction deformation we construct a family of quantum 
Hamiltonians whose ground states interpolate between the toric code 
wavefunction~\eqref{CTT_eq: toric code GS} and the fully polarized state 
$\ket{\psi_p} = \bigotimes^{\ }_{i} \vert \sigma^{z}_{i}=+1 \rangle$. 
This can be accomplished by using ``stochastic matrix form" (SMF) 
decompositions~\cite{CTT_Castelnovo2005} for a generic wavefunction of 
the form
\begin{equation} 
\ket{\psi} 
\propto 
\sum_{\{\bold{s}\,:\, \phi_p(\bold{s})=+1 \,\, \forall p\}} 
\exp\left(\frac{h}{2} E_{\bold{s}}\right) \:
c_{\bold{s}} \ket{\bold{s}} 
. 
\label{CTT_eq: SMF GS}
\end{equation}
For $h=0$ this is the toric code ground state, while 
for $h \to \infty$ the polarized state is exponentially selected, 
provided that $E_{\bold{s}}$ is `sufficiently peaked' at $\bold{s} = \{z_j = +1\}$. 
The corresponding SMF Hamiltonian can simply be expressed as a sum of projectors that 
annihilate~\eqref{CTT_eq: SMF GS}. So long as the function(al) $E_{\bold{s}}$ 
can be written as a sum of local terms in the variables $\{z_j\}$, then the 
Hamiltonian can be constructed as a sum of local operators expressed in terms of 
the Pauli matrices $\sigma^{x,y,z}_j$. 

The simplest example is given by the choice $E_{\bold{s}} = \sum_j z_j$. 
It is straightforward to verify that
\bea
H 
&=& 
- 
J_{e} 
\sum^{\ }_{s} 
    A^{\ }_{s} 
+ 
J_{e}
\sum^{\ }_{s} 
  \exp{\left({- h \sum^{\ }_{i \in s} \sigma^{z}_{i}} \right)}
\label{CTT_eq: Kitaev field Ham 1}
\eea
is a sum of projectors, each of which 
annihilates~\eqref{CTT_eq: SMF GS}~\cite{CTT_Castelnovo2005,CTT_Castelnovo2008_fK}.\footnote{A 
generic construction scheme for SMF Hamiltonians given arbitrary $E_{\bold{s}}$ functionals 
is provided in Ref.~\onlinecite{CTT_Castelnovo2005}.} 
However, this remains true if the summation in~\eqref{CTT_eq: SMF GS} is generalized 
from $\{\bold{s}\,:\, \phi_p(\bold{s})=+1 \,\, \forall p\}$ to 
$\{\bold{s}\,:\, \phi_p(\bold{s})=\overline{\phi}_p \,\, \forall p\}$, for any configuration 
$\{\overline{\phi}_p\}$ of the plaquette eigenvalues. 
In order to arrive at the desired result, we need to add an appropriate energy cost 
$- J_{m} \sum^{\ }_{p} B^{\ }_{p}$ to the unwanted ground 
states~\cite{CTT_Castelnovo2008_fK}, 
\bea
H 
&=& 
- 
J_{m} 
  \sum^{\ }_{p} 
    B^{\ }_{p} 
- 
J_{e} 
\sum^{\ }_{s} 
    A^{\ }_{s} 
+ 
J_{e}
\sum^{\ }_{s} 
  \exp{\left({- h \sum^{\ }_{i \in s} \sigma^{z}_{i}} \right)}
\label{CTT_eq: Kitaev field Ham}
\eea

Not surprisingly, for $h=0$, the system reduces to the toric code 
Hamiltonian \eqref{CTT_Eq:ToricCode}, up to a trivial constant shift in energy. 
Moreover, in the limit $\vert h \vert \ll 1$ one can expand the exponential in 
Eq.~\eqref{CTT_eq: Kitaev field Ham}, and to first order one obtains precisely the 
Hamiltonian deformation studied in the previous section 
with a longitudinal field $h_z = 2 h J_e$. 
This equivalence is eventually lost for larger values of $h$, albeit by 
construction both models favor indeed the same fully polarized, paramagnetic state 
for large fields $h \gg 1$. 

For convenience of notation, in the following we shall replace the sum over 
$\{\bold{s}\,:\, \phi_p(\bold{s})=+1 \,\, \forall p\}$ in 
Eq.~\eqref{CTT_eq: SMF GS} in terms of the (Abelian) group $G$ 
of all spin flip operations obtained as products of star operators 
$A_s$~\cite{CTT_Kitaev2003,CTT_Hamma2005}, so that the ground state wavefunction of 
model \eqref{CTT_eq: Kitaev field Ham} reads 
\bea
\vert \psi_0^{(\alpha)} \rangle 
&=& 
  \frac{1}{\sqrt{Z^{\ }_{\alpha}}} 
  \sum^{\ }_{g \in G} 
    \exp{\left(\frac{h}{2} \sum^{\ }_{i} \sigma^{z}_{i}(g,\alpha) \right)}
      g \, \vert \Psi^{\ }_{\alpha} \rangle \,,
\label{CTT_eq: Kitaev field GS}
\\ 
{\rm with} \quad Z^{\ }_{\alpha} 
&=& 
\sum^{\ }_{g \in G} 
  \exp{\left(h \sum^{\ }_{i} \sigma^{z}_{i}(g,\alpha)\right)} 
\quad
{\rm and}
\quad
\sigma^z_i(g,\alpha) 
\equiv 
\langle \Psi^{\ }_{\alpha} \vert g \,
\sigma^{z}_{i}
\, g \vert \Psi^{\ }_{\alpha} \rangle 
. 
\nonumber 
\eea
Here the index $\alpha$ labels the $4$ topological sectors, i.e., the 
orbits of the action of the star operators. 
The eigenstate $\vert \Psi^{\ }_{\alpha} \rangle$ can be chosen arbitrarily amongst 
the elements of the orbit $\alpha$, since the action of the group $G$ on any such 
state generates the entire orbit. 
%
%

\subsubsection{Dimensionality reduction and the 2D Ising model} 
\label{CTT_sec: thermodynamic transition}

For convenience, we study the quantum phase transition to the paramagnetic 
state in the topological sector containing the fully polarized state 
$\ket{\psi_p} = \bigotimes^{\ }_{i} \vert \sigma^{z}_{i}=+1 \rangle$. 
The explicit form of the wavefunction reduces to
\beq
\vert \psi_0 \rangle 
= 
\frac{1}{\sqrt{Z}} 
  \sum^{\ }_{g \in G} 
    \exp{\left({h \sum^{\ }_{i} \sigma^{z}_{i}(g) / 2}\right)} \: 
      g \, \ket{\psi_p}
\quad
{\rm with}
\quad
\sigma^{z}_{i}(g) 
\equiv 
\bra{\psi_p} g \,\sigma^{z}_{i}\, g \ket{\psi_p}\,.
\label{CTT_eq: Kitaev field GS sector 0}
\eeq

A generic configuration $g\ket{\psi_p}$ is uniquely specified by the 
set of star operators acting on the reference configuration $\ket{\psi_p}$, 
modulo the action of the product of all the star operators (which is equal to the identity). 
Thus, there is a $1$-to-$2$ mapping between $G = \{ g \}$ and the 
configuration space $\Theta = \{ \bftheta \}$ of an Ising model with 
classical degrees of freedom $\theta^{\ }_{s}$ living on the sites $s$ of the square 
lattice, where for example $\theta^{\ }_{s} = -1$ ($+1$) means that the 
corresponding star operator is (not) acting in the associated $g$. 
Since each $\sigma$-spin can be flipped only by its two neighboring 
$\theta$-spins, then 
$
\sigma^{\ }_{i} 
\equiv 
\theta^{\ }_{s}\theta^{\ }_{s^{\prime}_{\ }}
$, 
where $i$ labels the bond between the two neighboring sites 
$\langle s,s^{\prime}_{\ } \rangle$. 
Using this mapping, the ground state wavefunction of the model, 
Eq.~\eqref{CTT_eq: Kitaev field GS sector 0}, can be rewritten as 
\beq
\vert \psi_0 \rangle 
= 
\frac{1}{\sqrt{Z}}
  \sum^{\ }_{\bftheta \in \Theta} 
    \exp{\left(\frac{h}{2} \sum^{\ }_{\langle s,s^{\prime}_{\ } \rangle} 
       \theta^{\ }_{s} \theta^{\ }_{s^{\prime}_{\ }} \right)} \:
      g(\bftheta) \, \ket{\psi_p} \,, 
\eeq
where 
$
Z 
= 
\sum^{\ }_{\bftheta \in \Theta} 
  e^{h \sum^{\ }_{\langle s,s^{\prime}_{\ } \rangle} 
     \theta^{\ }_{s} \theta^{\ }_{s^{\prime}_{\ }}}
$. 

All equal-time correlation functions of the ground state that can be 
expressed in terms of the classical $\theta$-spins can thus be 
computed as correlation functions of a $2D$ classical Ising model with 
reduced nearest-neighbor coupling $J/T = h$~\cite{CTT_Castelnovo2005}. 
For small values of the field $h$, we can identify the equal-time correlation 
functions of the quantum model in the topological phase with the correlation
functions of the high-temperature disordered phase of the 2D Ising model.
Vice versa, the large field paramagnetic state of the quantum model 
exhibits equal-time correlators that match those of the low-temperature 
Ising ferromagnetically ordered state.
For an intermediate magnetic field strength $h_c$ the ground-state wavefunction 
of the quantum system becomes ``critical" and scale invariant, precisely when 
the corresponding classical spin system undergoes its thermal phase transition at 
$h^{\ }_{c} = (1/2) \ln (\sqrt{2} + 1) \simeq 0.441$.

While the thermal phase transition in the 2D Ising model can be described in
terms of a local order parameter, e.g. its magnetization, we note that this is not
the case for the original quantum model. For instance, the magnetization in the original 
$\sigma$-spin language translates into the nearest-neighbor spin-spin correlation, 
i.e., the energy, in the $\theta$-spin language, 
\bea
m(h) 
= \! 
\frac{1}{N}\sum_i 
  \langle \psi_0 \vert \sigma^{z}_{i} \vert \psi_0 \rangle 
= \! 
\sum^{\ }_{\bftheta \in \Theta} 
  \frac{e^{h \sum^{\ }_{\langle s,s^{\prime}_{\ } \rangle} 
     \theta^{\ }_{s} \theta^{\ }_{s^{\prime}_{\ }}}}{Z} \!\!
    \left[ 
      \frac{1}{N} \!\! \sum^{\ }_{\langle s,s^{\prime}_{\ } \rangle}
        \theta^{\ }_{s}\theta^{\ }_{s^{\prime}_{\ }} \! 
    \right] \!
= \!
\frac{1}{N} \,E_{\rm Ising}(h)
\;, 
\nonumber
\eea
and one concludes that $m(h)$ does not vanish on either side of the transition. 
It is continuous 
across the transition, but there is a singularity in its first derivative 
\beq 
\frac{\partial m}{\partial h} 
= 
\frac{1}{N}\,\frac{\partial E_{\rm Ising}}{\partial h}
= 
-h^2\;\frac{1}{N}\,C_{\rm Ising}(h) 
\;, 
\eeq 
as the heat capacity $C_{\rm Ising}(h)$ of the classical Ising model diverges 
logarithmically at $h_c$. 
Thus, while the original quantum model does not allow to define a local order 
parameter for the phase transition, there is nevertheless a signature of this transition
in terms of a singularity in the derivative of a local observable. 


\subsubsection{Topological entropy} 
\label{CTT_sec: topo entropy}

We have seen that the unsual properties of topologically ordered phases are inherently 
linked to non-local properties and a peculiar type of long-range entanglement. 
With all local correlations being short ranged and non-local correlation functions 
being tedious to define in more general examples, 
it is helpful to turn to entanglement measures and to define
a {\em topological entropy} that allows to 
detect and characterize topological phases~\cite{CTT_Levin2006,CTT_Kitaev2006}.

A common measure of quantum entanglement is given by the von Neumann entropy, 
$\SvN = -\Tr[\rho\ln\rho]$, where $\rho$ is the density  matrix of the system. 
Typically one considers a (smooth) bipartition $(\A,\B)$ of the system 
$\S = \A \cup \B$, with boundary size $L$ 
(which in two dimensions is simply its length), 
and computes the reduced density matrix  $\rho_\A = \Tr_B \rho$. 
The von Neumann entropy $\SvN(\A) = -\Tr[\rho_\A\ln\rho_\A]$ is also known as 
the (bipartite) entanglement entropy, and reflects the `amount of information' 
shared between the two subsystems across the boundary. 
In ordinary short-range correlated phases, the nature of subsystem $\B$ 
influences subsystem $\A$ only up to some finite distance (of the order of 
the correlation length) away from the boundary, and vice versa. 
To leading order in $L$ the van Neuman entropy thus takes a form 
$\SvN(\A) = \alpha L + \ldots$, where the coefficient $\alpha$ is non-universal. 

\begin{figure}[t]
         \begin{center}
         \includegraphics[width=0.98\columnwidth]{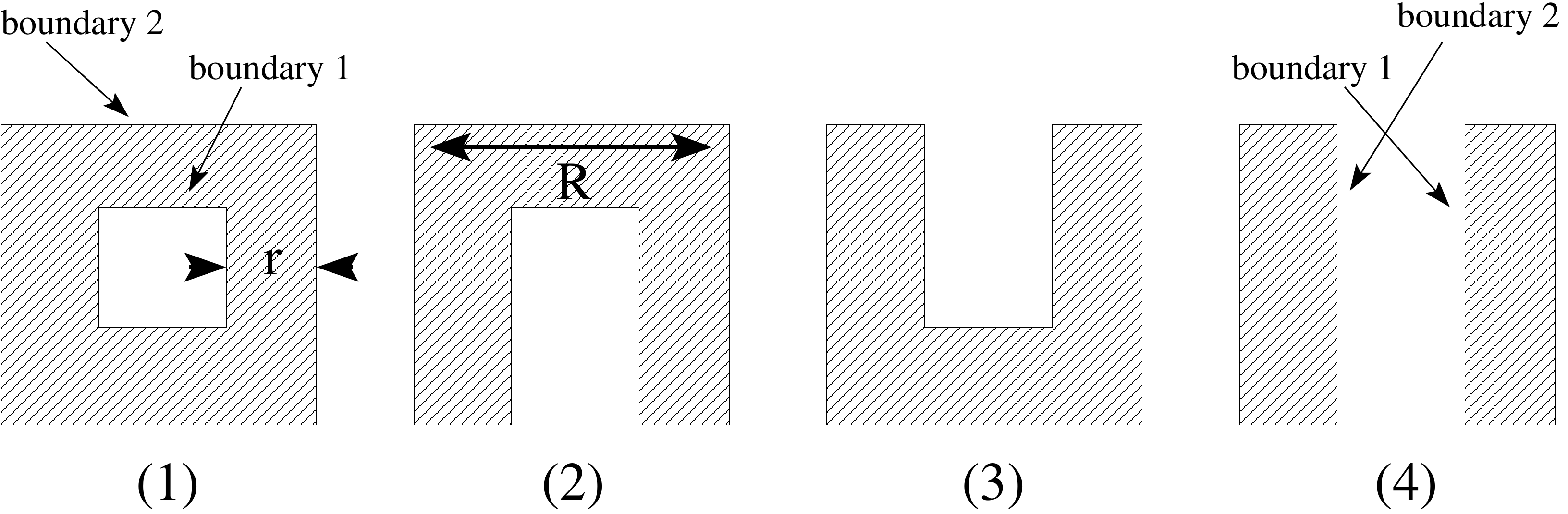}
	\end{center}
          \caption{{\em Bipartition scheme for the topological entropy:} 
	           Illustration of the four distinct bipartitions $\S=\A_i \cup \B_i$ used to 
                   compute the topological entropy in Ref.~\onlinecite{CTT_Levin2006}. 
                   The shaded area represents subsystem $\A_i$.}
\label{CTT_fig: topological partitions}
\end{figure}

If, on the other hand, non-local information is stored across a boundary as in the case of a 
topologically ordered state, 
it must show up as a further contribution to the entanglement 
between the two bipartitions $\A$ and $\B$. 
It was argued in Refs.~\onlinecite{CTT_Hamma2005,CTT_Levin2006,CTT_Kitaev2006} that this contribution 
assumes the form of a 
correction of order one to the previous scaling form, e.g.
\bea
\SvN(\A) = \alpha L - \gamma + O(1/L) 
, 
\label{CTT_eq: SvN scaling} 
\eea
with all other subleading terms vanishing in the limit $L \to \infty$. 
This correction $\gamma \geq 0$ is found to be {\em universal}, and it is related 
to the so-called ``total quantum dimension" of the underlying topological 
phase~\cite{CTT_Levin2006,CTT_Kitaev2006}. 
A particularly elegant way to calculate the universal contribution $\gamma$ is to
carefully choose an appropriate set of different bipartitions, 
such that a linear combination of the respective entanglement entropies cancels out 
all non-universal terms 
(at least in the limit of the correlation length being much smaller than 
the size of the bipartitions). 
Adopting the scheme proposed in Ref.~\onlinecite{CTT_Levin2006} and 
illustrated in Fig.~\ref{CTT_fig: topological partitions}, 
we define the topological entropy as: 
\beq
\Stopo
= 
\lim^{\ }_{r, R \to \infty} 
\left[\vphantom{\sum} 
  - \SvN{(\A_1)} 
  + \SvN{(\A_2)} 
  + \SvN{(\A_3)} 
  - \SvN{(\A_4)} 
\right] 
, 
\label{CTT_eq: topo entropy}
\eeq
where the indices $i=1,\ldots,4$ label the four different bipartitions 
$\S=\A_i \cup \B_i$. 
Note that bipartitions $(\A_1,\B_1)$ and $(\A_4,\B_4)$ combined have the same 
total boundary, with the same number and type of corners, as bipartitions 
$(\A_2,\B_2)$ and $(\A_3,\B_3)$ combined. 
As a result, the non-universal part of the entanglement entropy due to local 
correlations cancels out in Eq.~\eqref{CTT_eq: topo entropy}. 
However, only bipartitions $(\A_1,\B_1)$ and $(\A_4,\B_4)$ are topologically 
non-trivial, and the universal correction to the corresponding entanglement 
entropies is doubled. 
This definition of $\Stopo$ therefore measures twice the value of $\gamma$, 
i.e., $\Stopo = 2 \gamma$.

Any approach to compute the topological entropy 
faces the challenge of isolating a subleading term of order one in a quantity, 
the entanglement entropy, that scales with the size $L$ of the bipartition.
(This, of courses, raises the question whether there are better measures of the
long-range entanglement of topological phases.)
Perturbative techniques can be used only if the achieved accuracy is greater
than $O(1/L)$. 
Likewise, it is rather difficult to obtain reliable numerical estimates~\cite{CTT_Furukawa2007}. 
On the other hand, at zero temperature the density matrix of the system becomes the 
projector onto the ground state $\ket{\psi_0}$, namely $\rho = \vert\psi_0\rangle\langle\psi_0\vert$, 
and one can attempt to calculate $\SvN$ and $\Stopo$ directly whenever 
an explicit expression for the ground state wavefunction is available. 
Hereafter, we follow this approach and compute the topological entropy for the 
``wavefunction deformation" of the toric code presented in Sec.~\ref{CTT_sec: model}.


\subsubsection{Topological entropy along the wavefunction deformation} 
\label{CTT_sec: topo entropy fK}

To calculate the topological entropy along the wavefunction deformation
we consider a smooth bipartition $(\A,\B)$ of the system. 
Let us define the subgroup $G^{\ }_{\A} \subset G$ that acts solely on $\A$ 
and leaves $\B$ invariant: 
$
G^{\ }_{\A} 
= 
\{ g=g_\A \otimes g_\B \in G \;\vert\; g^{\ }_{\B} = \openone^{\ }_{\B} \}
$. 
Likewise we define $G^{\ }_{\B}$. 
Following Ref.~\cite{CTT_Castelnovo2008_fK} the von Neumann entropy $\SvN(\A)$ 
from the ground state wavefunction in Eq.~\eqref{CTT_eq: Kitaev field GS sector 0} 
can be expressed as
\bea
\SvN(A) 
&=& 
- \frac{1}{Z} \sum^{\ }_{g \in G} e^{-h E^{\ }_{g}} \,\,
    \ln 
      \left(
        \frac{1}{Z}\sum^{\ }_{f \in G^{\ }_{A},\, k \in G^{\ }_{B}} 
              e^{-h E^{\ }_{f\! g k}}
      \right) \,,
\label{CTT_eq: 1-body von Neumann entropy}
\eea
where $E_g = -\sum_i \sigma^z_i(g)$. 
This is the lattice equivalent of the von Neumann entropy 
obtained by Fradkin and Moore for quantum Lifshitz field 
theories~\onlinecite{CTT_Fradkin2006}. 
Note that this expression for the von Neumann entropy can also be interpreted 
as the ``entropy of mixing" (or configurational entropy) of the allowed boundary 
configurations in $G$
(this was first shown in Ref.~\cite{CTT_Castelnovo2008_fK}, 
and later investigated in more detail in Ref.~\cite{CTT_Stephan2009}).

Employing the change of variables to classical Ising degrees of freedom $\{\theta_s\}$ 
(see Sec.~\ref{CTT_sec: thermodynamic transition}), and omitting the 
technical details~\cite{CTT_Castelnovo2008_fK} required to deal with configurations 
of the form  $f\!gk$ ($f \in G^{\ }_{A}$ and $k \in G^{\ }_{B}$), 
one can use Eq.~\eqref{CTT_eq: 1-body von Neumann entropy} to obtain the 
topological entropy of the system as a function of $h$ (with the bipartition 
scheme illustrated in Fig.~\ref{CTT_fig: topological partitions}) 
\bea
\Stopo 
&=&
\lim^{\ }_{r, R \to \infty} 
\left\{\vphantom{\sum} 
  \frac{\sum^{\ }_{g \in G} e^{-h E_g}}{Z} 
\right. 
\nonumber\\
&&
\;\;
\times 
\left. 
  \ln 
    \frac{\left[ Z^{\partial}_{1}(g) + 
                 Z^{\partial,\,\textrm{twisted}}_{1}(g) \right]
          \left[ Z^{\partial}_{4}(g) + 
	         Z^{\partial,\,\textrm{twisted}}_{4}(g) \right]}
         {Z^{\partial}_{2}(g) Z^{\partial}_{3}(g)}
\vphantom{\sum^{\ }_{g\in G}}\right\} 
. 
\label{CTT_eq: formula for the topo entropy}
\eea
Here $Z^{\partial}_{2}(g)$ and $Z^{\partial}_{3}(g)$ represent the partition 
functions of an Ising model with nearest-neighbor interactions of reduced 
strength $J/T = h$, and with fixed spins along the boundary of bipartitions 
$2$ and $3$, respectively. 
Likewise, $Z^{\partial}_{1,4}(g)$ are analogous partition functions for 
bipartitions $1$ and $4$, respectively. 
Note that the boundaries in bipartitions $1$ and $4$ have two disconnected 
components, labeled `boundary $1$' and `boundary $2$' in 
Fig.~\ref{CTT_fig: topological partitions}. 
The partition functions $Z^{\partial,\,\textrm{twisted}}_{1,4}(g)$ differ from 
$Z^{\partial}_{1,4}(g)$ in that all the (fixed) spins belonging to boundary 
$2$ in bipartitions $1$ and $4$, respectively, have been flipped. 

The sum over $g$ in Eq.~\eqref{CTT_eq: formula for the topo entropy} acts 
as a weighed average of the logarithmic term over all possible values of the 
spins at the boundary. Notice that the partitions 
with two boundaries, and hence with non-trivial topology, are those that 
appear with two contributions (bipartitions $1$ and $4$), corresponding to 
the sum of each boundary condition with its twisted counterpart. 

In the topological phase (i.e., in the disordered phase of the corresponding 
Ising model), where correlations are short ranged, the choice of boundary 
conditions affects the partition function of the system only with exponentially 
small corrections. 
Thus, we can expect to have 
$
Z^{\partial}_{1}(g)Z^{\partial}_{4}(g) 
\simeq 
Z^{\partial,\,\textrm{twisted}}_{1}(g)Z^{\partial}_{4}(g) 
\simeq 
\,\ldots\, 
\simeq 
Z^{\partial}_{2}(g) Z^{\partial}_{3}(g)
$
and the topological entropy becomes $\Stopo = \ln4$. 
On the other hand, in the polarized paramagnetic phase 
(i.e., in the ferromagnetically ordered phase of the Ising model) 
the partition function of a system with twisted boundary conditions is 
exponentially suppressed with respect to the one without the twist. 
Thus, 
$
Z^{\partial}_{1}(g) 
\gg 
Z^{\partial,\,\textrm{twisted}}_{1}(g) 
$, 
$
Z^{\partial}_{4}(g) 
\gg 
Z^{\partial,\,\textrm{twisted}}_{4}(g) 
$, 
while 
$
Z^{\partial}_{1}(g)Z^{\partial}_{4}(g) 
\simeq 
Z^{\partial}_{2}(g) Z^{\partial}_{3}(g)
$ 
still holds. 
This leads to $\Stopo = 0$. 

Using a high-temperature expansion for $h < h_c$, and appropriate 
Ising duality relations for $h > h_c$, 
one can show that the behavior of the topological entropy across the 
transition is {\em discontinuous}, with a sudden jump from 
$\Stopo = \ln4$ to $\Stopo = 0$ at $h_c$, in spite of the otherwise 
continuous nature of the transition~\cite{CTT_Castelnovo2008_fK}. 
This discontinuity of the topological entropy is probably less surprising 
if one keeps in mind that the topological entropy is inherently linked to the
quantum dimensions of the elementary excitations of a phase~\cite{CTT_Levin2006,CTT_Kitaev2006}, 
and thus it must remain constant for the full extent of the phase.\footnote{
A similar behavior of the topological entropy has been found in numerical simulations \cite{CTT_Hamma2008}
of the Hamiltonian deformation discussed in section \ref{CTT_sec: longitudinal field}.}

%
%

\section{Thermal transitions}
\label{CTT_sec: thermal transitions} 

Our discussion so far was from a purely zero temperature perspective
of phase transitions involving topologically ordered phases. 
In conventional quantum phase transitions -- where a local order parameter 
acquires a non-vanishing expectation value -- the zero-temperature quantum 
critical point controls an extended portion of the finite temperature phase 
diagram, the so-called ``quantum critical region'' or ``quantum critical fan". 
One thus wonders to what extent a similar behavior can be expected to 
hold for quantum phase transitions involving topological phases of matter. 
To address this question, we first discuss a framework to measure
and characterize topological order at finite temperature, using both 
non-local correlators and a generalization of the topological entropy.
This will further allow to discuss when topological 
order survives in the presence of thermal fluctuations, 
and under what conditions one can expect to observe a 
{\em finite-temperature} phase transition. 

\subsection{Non-local order parameters at finite temperature}
\label{CTT_sec: non-local ops at finite T}

At zero temperature we can typically characterize topological order by non-local 
order parameters, for instance the winding loop operators in the 
toric code~\cite{CTT_Kitaev2003}. 
If we want to use the same order parameters at {\em finite} temperature,
one needs to carefully consider the role of defects, e.g. a flipped spin occurring 
along the path of a winding loop operator, which changes its expectation value.
In a topological phase with excitation gap $\Delta$, a finite temperature $T$ 
induces a finite density $\rho \sim e^{-\Delta / T}$ of defects. 
Our ability to `detect' the topological phase using loop operators relies on 
the ability to find loops that do not encounter any thermal defects along 
their path. 
However, the probability for the occurrence of such unaffected loops scales as 
$(1-\rho)^\ell$, where $\ell$ is the total length of the loop. 
For winding loops with $\ell \geq L$, where $L$ is the linear size of the system,
this probability vanishes exponentially fast in the thermodynamic limit --
irrespective of how low the temperature is.
This phenomenon was first discussed in the context of topological phases in 
Ref.~\onlinecite{CTT_Nussinov2006-7} and termed ``thermal fragility''.\footnote{
The  reader familiar with the connection between toric code and $\mathbb{Z}_2$ 
lattice gauge theory may recognize that this statement follows from the 
property that the expectation value of Wilson loop operators always vanishes in the 
limit of infinite loop size~\cite{CTT_Wegner1971}.}
This is quite distinct from conventionally ordered phases, 
where the thermal uncertainty in the expectation value of a local order parameter
scales with the density of defects, and becomes negligibly small at low temperature. 

Note that this thermal fragility of non-local order parameters cannot directly be equated to
a thermal instability of the topological phase.
The latter is determined by the nature of the anyonic defects, and stable topological 
phases at finite temperature are indeed possible \cite{CTT_Dennis2002}, which can be 
captured by the behavior of the topological entropy of the system \cite{CTT_Castelnovo2008_3d}.

For a {\em finite} system one can nevertheless define a {\em crossover temperature}
$T^*$. The free energy of a pair of defects, each with gap $\Delta$, is 
$F=2\Delta-T\ln N(N-1)\sim 2\Delta - 2T\ln N$. Above a temperature $T^*=\Delta/\ln N$ 
we always encounter excitations, but for $T\ll T^* \sim 1/\ln N$ the system is 
protected. This logarithmic reduction of the gap scale is no major 
technological problem for potential future devices built on such phases that might 
operate on system sizes $N\sim100$. It is, however, an interesting and important aspect 
when discussing topological stability.


\subsection{Topological entropy at finite temperature}
\label{CTT_sec: toric code finite T topo entropy}


To characterize the stability of a topological phase at finite temperature we now turn to 
topological entropy. 
While entanglement entropies are often considered zero-temperature quantities, 
we can readily generalize their definition in terms of the density matrix to finite 
temperature by simply considering $\rho(T) = \exp(-\beta H)/\Tr[\exp(-\beta H)]$,
where $H$ is the Hamiltonian and $\beta = 1/k_B T$. 

Computing the so-generalized topological entropy $\Stopo$ at finite 
temperature often turns out to be a significant challenge, even for state of the 
art numerical techniques~\cite{CTT_Furukawa2007}. 
On the other hand, some instances such as the toric code model are simple
enough to allow for a rigorous calculation as a function of both system size 
and temperature~\cite{CTT_Castelnovo2007_fTK,CTT_Castelnovo2008_3d,CTT_Castelnovo2007_cl}, 
as we briefly discuss in the following. 

In the toric code Hamiltonian \eqref{CTT_Eq:ToricCode} we have 
seen that  plaquette (P) and star (S) operators commute with one another. 
When calculating the finite-temperature generalizations of the entanglement and 
topological entropies, the resulting expressions are additive in 
these two contributions 
\bea
   S(\A;T) = S^{(P)}(\A;T/J_m) + S^{(S)}(\A;T/J_e) \,.
\eea
Therefore, vortex defects, i.e. plaquette operators with negative eigenvalues, 
are immaterial to the star contribution and vice versa. 
As a consequence, one can conveniently consider the two contributions separately.
Moreover, the system is symmetric upon exhanging $x$ with $z$ spin 
components, and plaquettes with stars. 
The two contributions $S^{(P)}$ and $S^{(S)}$ can thus be cast in the exact 
same analytical form. 

\begin{figure}[t]
    \begin{center}
     \includegraphics[width=0.75\columnwidth]{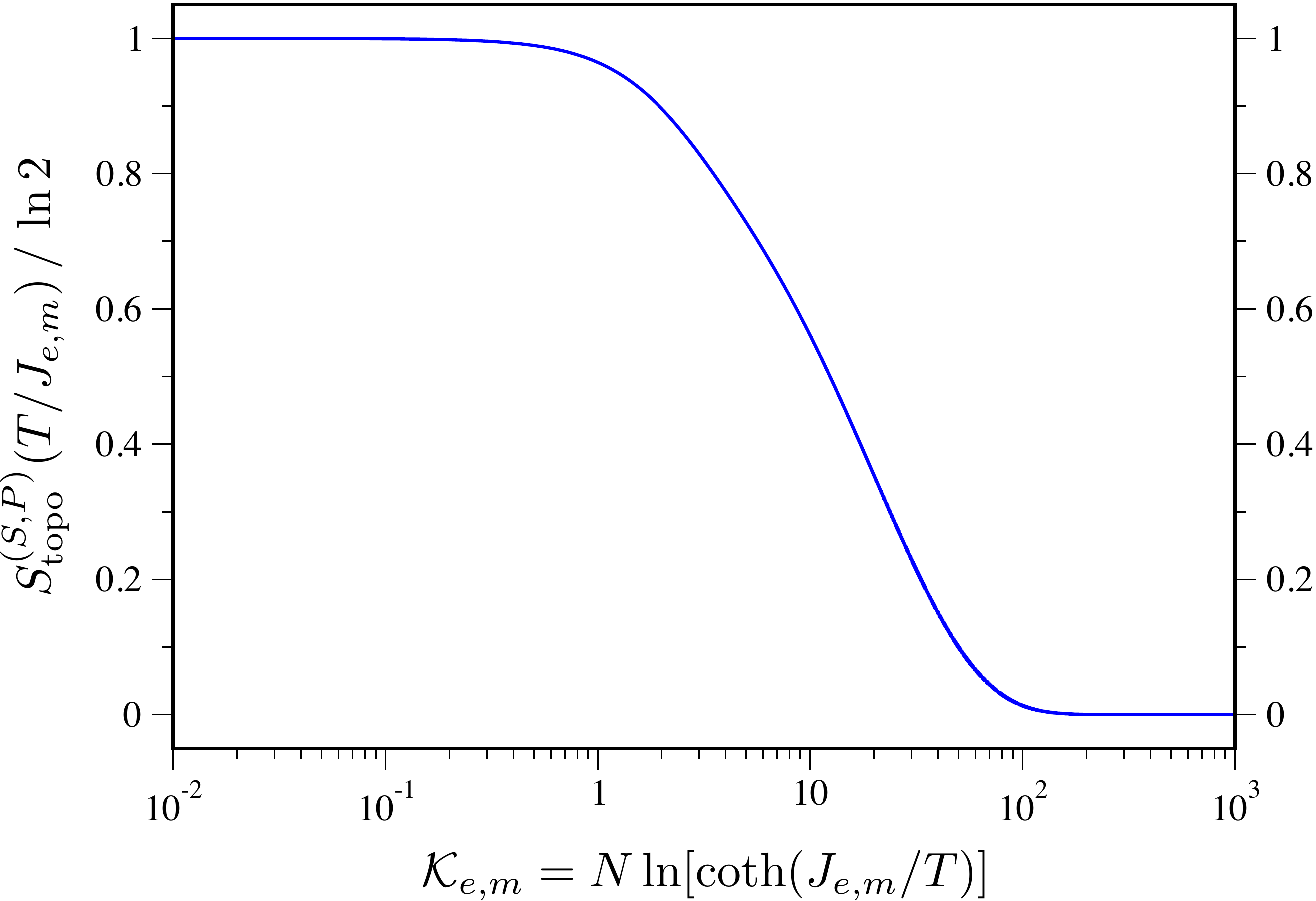}
     \end{center}
      \caption{{\em Finite-temperature topological entropy of the toric code:} 
                     $\Stopo^{(S,P)}(T/J_{e,m})$ as a function of $\K_{e,m} = N \ln[\coth(J_{e,m}/T)]$. 
}
\label{CTT_fig: Stopo 2D collapse}
\end{figure}

Deriving the explicit expressions for $\Stopo^{(P)}(T/J_m)$ and $\Stopo^{(S)}(T/J_e)$
as a function of temperature and system size $N$ is a rather cumbersome exercise 
and we refer to Ref.~\onlinecite{CTT_Castelnovo2007_fTK} for a detailed description. 
In both cases temperature and system size enter these expressions through the
product $\K_{e,m} = N \ln[\coth(J_{e,m}/T)]$, respectively.
Hence, the thermodynamic limit, $N \to \infty$, and the zero 
temperature limit, $T \to 0$, do not commute. 
The zero-temperature value of the topological entropy, $\Stopo=\ln4$, is recovered 
\emph{only} if the temperature is lowered to zero whilst keeping the system size finite. 
Vice versa, in the thermodynamic limit the topological 
entropy vanishes at any finite $T$, no matter how small. 

In Fig.~\ref{CTT_fig: Stopo 2D collapse} we plot the analytical expressions for the 
topological entropies $\Stopo^{(P)}(T/J_m)$ and $\Stopo^{(S)}(T/J_e)$ as a function 
of $\K_{e,m}$. 
For a fixed system size $N$, the plotted behavior indicates that the topological 
entropy decays to zero at a finite temperature. With $T \ll J_{e,m}$ we can expand 
$N \ln[\coth(J_{e,m}/T)] \sim 2N \exp{\left(-2J_{e,m}/T\right)}$, 
where the exponential corresponds to the Boltzmann weight for a single spin flip 
to occur in the ground state of the system, i.e., the density of defects. 
The decay in Fig.~\ref{CTT_fig: Stopo 2D collapse} occurs when 
$N \exp{\left(-2J_{e,m}/T\right)} \sim 1$, that is when the average \emph{number} 
of defect pairs in the system is of order unity. 
This translates into a crossover temperature $T^* \sim 2J_{e,m}/\ln N$, 
in agreement with the behavior of the non-local order parameters discussed
earlier.
%
%

\subsection{Fragile vs robust behavior: a matter of (de)confinement}
\label{CTT_sec: fragile vs robust}

The results in Sec.~\ref{CTT_sec: toric code finite T topo entropy} and 
Sec.~\ref{CTT_sec: non-local ops at finite T} show that the topological order 
in the 2D toric code is truly fragile to thermal fluctuations, 
in the sense that it is destroyed by a finite \emph{number} of defects. 
In the thermodynamic limit there is no finite-temperature phase transition.

{}From the behavior of the non-local operators one might \emph{mistakenly}
conclude that this fragility to thermal fluctuations is intrinsic to topological 
order~\cite{CTT_Nussinov2006-7}. 
On the other hand, we can compare this to the well-known example of a classical 
$Z_2$ lattice gauge theory in three dimensions~\cite{CTT_Kogut1979,CTT_gauge_refs}, which also 
lacks a local order parameter in the zero temperature limit. 
There, the $T=0$ state does not subside immediately to thermal fluctuations, 
and a finite temperature phase transition exists between 
a low-temperature phase with confined defects, and a high-temperature
phase where the defects are deconfined (see also Sec.~\ref{CTT_sec: longitudinal field}). 
Whereas non-local operators (such as Wilson loops around the whole system) 
vanish in both phases in the thermodynamic limit, the finite temperature phase 
transition is ultimately captured by the way that the winding loop expectation 
values vanish as the system size is increased 
(the area vs. perimeter law~\cite{CTT_Wegner1971}). 

Key to the existence of a finite temperature phase transition 
in the $Z_2$ lattice gauge theory is the confined nature of the thermal excitations 
in the low-temperature phase. 
Consistently, the deconfined nature of the defects in the topological phase of 
the toric code in two dimensions leads to the observed fragile behavior at 
finite temperature. 
If, on the other hand, we can devise a model where the excitations are confined, 
then we might expect a qualitatively different behavior. 
One example of the latter is the three-dimensional generalization of the toric
code~\onlinecite{CTT_Hamma2005_3dK}. 
In this model, plaquette and star operators are no longer dual to each other, and
while the star defects give rise to point-like excitations -- much like in the 2D case --
plaquette defects become objects which define closed loops (through plaquettes where 
the stabilizer condition of the ground states is violated), whose energy cost scales 
with the length of the loop. 
As a consequence, the plaquette defects do not fractionalize, but rather form confined 
structures whose characteristic size is controlled by temperature~\cite{CTT_Dennis2002}. 
This low-temperature phase with confined plaquette excitations is therefore stable to 
thermal fluctuations resulting in a finite-temperature (continuous) phase transition 
at $T_c / J_{m,e} \simeq 1.313346(3)$~\cite{CTT_Castelnovo2008_3d}. 
While the star contribution to the topological entropy exhibits a fragile behavior
(just as in the two-dimensional case), the plaquette contribution survives unaltered
up to the finite-temperature phase transition~\cite{CTT_Castelnovo2008_3d}. 

These results illustrate that the robustness of a zero-temperature 
topological phase depends crucially on the confined versus deconfined 
nature of the thermal excitations. Only if {\em all} excitations are confined, we 
expect quantum topological order to survive up to a finite temperature phase 
transition \cite{CTT_Dennis2002}. 

%
%

\section{Outlook}
\label{CTT_sec: outlook}

In this chapter we discussed phase transitions involving topologically 
ordered phases of matter in deformations of the toric code model. 
We focused on continuous quantum critical points 
where appropriate local correlators exhibit a divergent behavior. 
We showed that the growing correlations characterizing the critical region 
can be understood in terms of `dual' degrees of freedom, which undergo 
a conventional quantum phase transition between a disordered phase 
(corresponding to the topologically ordered one in the original system), and 
an ordered one characterized by a local order parameter (in the dual degrees 
of freedom). 
In this dual language, the underlying topological order is typically 
invisible -- the topological degeneracy being ``mapped out'' by the 
many-to-one correspondence in the definition of the dual degrees of 
freedom 
(see in particular the discussion in Sec.~\ref{CTT_sec: thermodynamic transition}
and Ref.~\cite{CTT_Feng2007}). 
Indeed, while the disruption of topological order across the transition 
occurs \emph{because of} the conventional phase transition in the underlying 
dual system, topological order survives unaffected up to the critical 
point (as whitnessed for instance by the step-function behavior of the topological 
entropy), irrespective of the continuous vs first-order nature of the 
transition (see Sec.~\ref{sec:otherdefomations}). 
The two phenomena -- the change in topological properties and the underlying 
conventional phase transition -- appear to be essentially unrelated in nature. 

One is then left to wonder whether it would be possible for a perturbation to 
cause topological order to subside or change 
\emph{without signature in any local correlators}. 
Such transitions could take place either between topologically ordered 
phases, or between a topologically ordered phase and a conventional 
paramagnet, somewhat akin to a glass transition in classical systems. 
Contrary to the examples considered in this chapter, these transitions
would be detected \emph{exclusively} by non-local observables. 
Or is it the case that even when there is no 
local order parameter on either side of the transition (whether in the 
original or dual degrees of freedom), there \emph{must} generically be 
a detectable singularity in high enough derivatives of some local observables? 
One approach that does not require any {\it a priori}  knowledge of an order parameter, 
and which has been argued to detect any singularities in local observables,
has recently been proposed by Zanardi {\it et al.} based on the information theoretic 
concept of fidelity~\cite{CTT_Zanardi2007}.
\footnote{
Examples of fidelity-based approaches to characterize quantum phase 
transitions involving topologically ordered phases of matter can be found 
for instance in Refs.~\onlinecite{CTT_Hamma2008,CTT_Eriksson2009}.} 
However, it remains to be seen whether fidelity-based techniques can be applied in 
systems where more conventional approaches fail.

\begin{figure}[t]
         \begin{center}
	\includegraphics[width=\columnwidth]{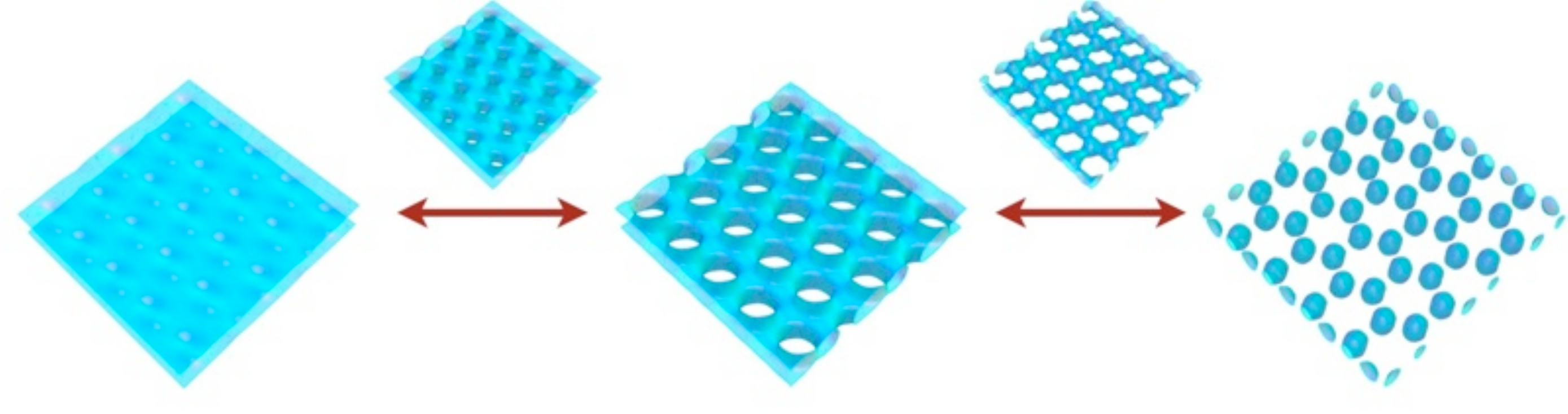}
	\end{center}
	\caption{{\em Topology driven quantum phase transition:}
               Two-dimensional surfaces with different topologies that are populated by anyonic 
                quantum liquids.
                A quantum phase transition driven by fluctuations of the surface topology 
                connects the anyonic liquid with topological order on two separated sheets (on the left) 
                and the anyonic liquid without topological order on decoupled spheres (on the right).}
               	\label{CTT_Fig:TopologyDrivenQPT}
\end{figure}

Another open question is whether the usual picture of a  ``fan''-like quantum critical region 
in the finite temperature phase diagram above a quantum critical point~\cite{CTT_CHN}
also applies to quantum phase transitions involving topological order. 
Does such a fan exist only in the presence of local order parameters, either in the system
at hand or a dual description, as was the case for the toric code and its dual description in 
terms of an Ising model? Or does a similar phenomenon appear at all continuous 
quantum phase transitions out of topological phases? 
A first attempt at addressing this question was recently made by Chung {\it et al.} in Ref.~\onlinecite{CTT_Chung2009}, where a quantum phase transition separating an Abelian from a non-Abelian topological phase was investigated in the context of an exactly solvable chiral spin liquid model.
What Chung {\it et al.} showed is how non-local order parameters can be used to define a crossover temperature 
$T^*$ -- akin to the one discussed in Sec.~\ref{CTT_sec: non-local ops at finite T} -- 
which thereby allows to define a temperature region reminiscent of the quantum critical fan in 
systems with local order parameters. 
The precise nature of this region, and whether its physics is 
as rich as for conventional 
quantum phase transitions, remains however an open question. 

More generally, we still need to identify a unifying theoretical framework that allows to 
describe topological phases and their phase transitions -- akin to the Landau-Ginzburg-Wilson
theory for conventional quantum phase transitions. 
A step towards such a more general description was taken in Ref.~\onlinecite{CTT_Kou2009} 
where the field-driven quantum phase transition in the toric code model was described in
terms of a mutual Chern-Simons Landau-Ginzburg theory. However, a generalization 
of this approach to non-Abelian phases is non-trivial. 

A unifying framework to describe quantum phase transitions for both Abelian and
non-Abelian topological phases has recently been proposed by some of us
by considering a general description in terms of quantum double models~\cite{CTT_Gils2009}. 
An Abelian or non-Abelian topological phase in a time-reversal invariant quantum lattice model, 
such as the toric code or the Levin-Wen model~\cite{CTT_Levin2005}, 
can be constructed from {\em chiral} topological quantum liquids populating the closed 
surface geometry obtained by ``fattening'' the edges of the lattice model, as illustrated for a 
honeycomb lattice in the middle panel of Fig.~\ref{CTT_Fig:TopologyDrivenQPT}. 
In this picture the quantum phase transition going from a topologically ordered phase to a 
(topologically trivial) paramagnetic state then corresponds to the transition between
different surface topologies as shown in Fig.~\ref{CTT_Fig:TopologyDrivenQPT}.
The topological phase corresponds to the limit of ``two sheets" illustrated on the left, 
while the paramagnet corresponds to the limit of ``decoupled spheres" illustrated on
the right. 
Plaquette flux excitations in the lattice model correspond to ``wormholes" in the 
quantum double model connecting the two sheets. Independent of their braiding statistics
it is the proliferation of these wormholes that drives a quantum phase transition between 
the two extremal states. 
The critical point between the topological and trivial phases is then described by a
``quantum foam" where the {\em surface topology fluctuates} on all length scales.
This concept of a ``topology-driven quantum phase transition" thereby provides a 
general framework that describes the quantum critical behavior of time-reversal 
invariant systems exhibiting topological order.

\section*{Acknowledgments}
S.T. and M.T. thank the Aspen Center for Physics for hospitality. 
C.C.'s work was supported by the EPSRC Grant No. GR/R83712/01 
and by the EPSRC Postdoctoral Research Fellowship EP/G049394/1. 
We thank E.~Ardonne, C.~Chamon, and E.~Kim for carefully reading the manuscript. 
C.C. is particularly grateful to C.~Chamon for stimulating discussions 
which led to some of the ideas presented in the Outlook section.


\section*{References}
\addcontentsline{toc}{chapter}{References}

%
%


\end{document}